\documentclass[%
 aip,
 amsmath,amssymb,
 reprint,%
]{revtex4-1}

\usepackage{graphicx}
\usepackage{dcolumn}
\usepackage{bm}

\usepackage[utf8]{inputenc}
\usepackage[T1]{fontenc}
\usepackage{mathptmx}
\usepackage{etoolbox}

\usepackage{xcolor}

\makeatletter
\def\@email#1#2{%
 \endgroup
 \patchcmd{\titleblock@produce}
  {\frontmatter@RRAPformat}
  {\frontmatter@RRAPformat{\produce@RRAP{*#1\href{mailto:#2}{#2}}}\frontmatter@RRAPformat}
  {}{}
}%
\makeatother
\begin{document}

\preprint{AIP/123-QED}

\title[Windmilling clusters of active quadrupoles]{Windmilling clusters of active quadrupoles}
\author{M. Rosenberg*}
 \email{rosenberg@thphy.uni-duesseldorf.de}
\affiliation{ Institut für Theoretische Physik II: Weiche Materie, Heinrich-Heine-Universit\"at D\"usseldorf, Universit\"atsstr. 1, D-40225 D\"usseldorf, Germany 
}%
\author{H. L\"owen}
\affiliation{ Institut für Theoretische Physik II: Weiche Materie, Heinrich-Heine-Universit\"at D\"usseldorf, Universit\"atsstr. 1, D-40225 D\"usseldorf, Germany 
}%

\date{\today}

\begin{abstract}
Active matter has thrived in recent years, driven both by the insight that it underlies fundamental processes in nature, and by its vast potential for applications. This allows for innovation both inspired by experimental observations, and by construction of novel systems with desired properties. In this paper, we develop a novel system in the search for a new kind of pattern formation: microstructural motifs with orthogonal alignment. Taking a simple active Brownian particle (ABP) model applied to dumbbell-shaped particles, we add a quadrupolar interaction by positioning two antiparallel magnetic dipolar moments on each particle. We find that the phase behavior is determined by the competition between active motion and the orthogonal alignment favored by quadrupolar attraction. By varying these quantities, we are able to tune both the internal structure of the aggregates, and find a surprising stability of triangular aggregates, to the point of clusters of size $N=3$ being strongly overrepresented. Although none of the component particles are chiral, the resulting structures spin in a random, fixed direction due to combination of the polarity of the active motion. This results in an ensemble of windmilling (randomly spinning in a circular motion) aggregates with windmill-like shape (due to the three- or four core component dumbbells). Ultimately, this simple model shows an interesting range of microstructural motifs, with great potential for experimental implementations.

\end{abstract}

\maketitle

\section{\label{sec:intro} Introduction}

Active matter consists of particles that take energy from their environment and channel this into propulsion, giving rise to many interesting dynamic and collective phenomena \cite{Gompper2020,Schmidt19,vrugt2025}. Often inspired by real-life creatures such as birds, one simple approach to refining this model is the inclusion of some form of local ordering, as in the Vicsek model\cite{Vicsek95,Negi2022,Liebchen2017,Broeker2023}. This can also be done by exploring the effects of different particle shapes\cite{Kraft13,Riedel2024}, such as dumbbells\cite{Schwarzendahl23,Clopes2022,Dennison2017}, ellipsoids\cite{Theers2018,Bickmann2022}, squares\cite{Prymidis2016} or even triangles\cite{Junot2022}.
Another promising avenue is to include longer-range interactions, such as magnetic dipoles interactions between the active particles. These have recently been studied in both 2-dimensional\cite{Liao2020,Othman2024,Sese-Sansa2022} and 3-dimensional active contexts\cite{Royall25, Kelidou24,Kaiser2020,Kaiser2021,Guzmán-Lastra2016,Klumpp19,Golestarian18,Guzmán-Lastra24,Koessel2019}, and in more complex geometries, often inspired by the diffusion of magnetotactic bacteria\cite{Klumpp25,Erdal18,Concha21}. The latter is of interest not only from a fundamental biological perspective, but also in order to drive the design of microscale magnetic machines or magnetic robots\cite{Jafari21,Aubret2018,Vincenti2019,Holm22,Valeriani23,Golestarian20,Opsomer25}. For the efficacy of such applications, understanding the self-assembly of these particles is critical, which has given rise to multiple investigations\cite{Ginot2018,Mallory18,Mallory2017,Hall20,Royall24,Erbe19,Spiteri17}.

With such ample examples, it is not surprising that a wide variety of states can be found in any phase diagram of such systems, with or without activity\cite{Lowen19,Sciortino12,Vandewalle24,Filion20,Mason2025,Gang22}. However, certain microstructural motifs remain elusive. In particular, the alignment between active anisotropic particles frequently favors the parallel, especially in magnetic dipolar systems where the minimum of the dipole-dipole energy is found in a head-to-tail configuration.
In this paper, we would like to access a more rich and varied microstructure by introducing a potential that, at its two-particle energetic minimum, would favor orthogonal alignment. This can easily be achieved by combining two antiparallel dipoles to create a quadrupole\cite{Nelson19}. Within the context of active particles, this yields an interesting competition as the most repulsive magnetic alignment in parallel would be favored by the activity for an elongated particle shape. By studying the phase diagram, we find the solution is more nuanced- specifically, we discover a new combined stable structure of triangles. Furthermore, due to the polar nature of the activity, these three-particle aggregates are only stable in the configuration which additionally allows them to spin in a circular motion. Larger aggregates also form, and as their interparticle orientational alignment at close range is dominated by the magnetic interactions, are not able to form sufficiently blocked aggregates: therefore these also rotate. As these circular, rotating motions resemble the rotation of windmills, we refer to them as {\it windmilling clusters}. This formation of a mixture of base-triangular and base-orthogonal motifs disrupts both the standard clustering models we would expect, as well as deviating from the expected orthogonal structure formation, but also does not follow any nucleation expected from active particles. Moreover, this is an  feasible system to implement in experiment. 

This manuscript is structured as follows. In section \ref{sec:methods}, we will present a more detailed overview of the model and the simulation techniques used. The subsequent section \ref{sec:results} is divided into three topics: in subsection \ref{subsec:passive}, we explore the implications of the quadrupolar model in absence of activity. Subsection \ref{subsec:phase} is dedicated to the phase diagram of magnetic dumbbells, and subsection \ref{subsec:micro} discusses the microstructure of the clustering in more detail to show which motifs are found. These results are summarized in section \ref{sec:concl}.

\section{\label{sec:methods} Methods}

\subsection{\label{subsec:model} Brownian Dynamics Simulations}

The data shown in this work were obtained by Brownian Dynamics simulations, more specifically by overdamping the Langevin translational and rotational equations of motion. Particles were confined to the $x,y$ plane and only permitted to rotate around the $z$-axis in order to create a strictly 2D system.  The system area $A$ and the number of dumbbells $N=1000$ are kept constant in each simulation run. Furthermore, the particles are in contact with a heat bath at temperature $T$. The area of the quadratic simulation box for each run was calculated based on the desired particle densities of $\phi = 0.05, 0.15, 0.3$. Since the particle surface area is $2 \cdot r^2 \pi = \frac{\pi}{2} \sigma^2$, this gives $\phi = \frac{\sigma^2 \pi N}{2 A}$, with $A = l^2$ for a box of length $l$. We now turn to defining the reduced simulation units. As the basis of our unit system, we fix the length scale $\sigma$, the diameter of each dumbbell component, the energy scale as the simulation temperature $k_BT$ and the relaxation time $D_{rot} = 0.1$. We then proceed to set $\sigma =1$, $k_BT = 1$ and $D_{rot} = 0.1$.
 For an arbitrary simulation particle, we can write the Langevin equations of motion as:
\begin{align*}
     m \; \mathbf{\dot{v}}(t) = \mathbf{F}(t) -v_o \hat{n} - \gamma_t \mathbf{v}(t) + \sqrt{2k_BT \gamma_t} \; \bm{\eta}_t(t) \\
     \mathbf{I} \; \mathbf{\dot{\bm{\omega}}}(t) = \bm{\tau}(t) - \gamma_r \bm{\omega}(t) + \sqrt{2 k_BT \gamma_r} \; \bm{\eta}_r(t)
\end{align*}
\noindent
where the superscript dot $\mathbf{\dot{}}$ represents the time derivative, $m$ (resp. $\mathbf{I}$) are the particle mass (inertia tensor), $\mathbf{v}$ (resp. $\bm{\omega}$) are the velocity (angular velocity), $\bm{F}$ (resp. $\bm{\tau}$) are the forces (torques) acting on the particle, $\gamma_t$ (resp. $\gamma_r$) are the translational (rotational) diffusion coefficients and $\bm{\zeta}_t$ (resp. $\bm{\zeta}_r$) are random forces (torques). These random forces and torques are chosen to be Gaussian white noise: each component has a mean of $0$ and a variance following the fluctuation-dissipation theorem $\langle \eta_t^{\alpha} (t), \eta_t^{\beta}(t') \rangle = \delta_{\alpha,\beta} \; \delta(t-t')$ where the angle brackets denote an ensemble average, and $\alpha,\beta$ denote spatial coordinates. The additional $-v_0 \hat{n}$ term represents the activity with velocity $v_0$ parallel to the particle director $\hat{n}$. In order to sample the Brownian Dynamics regime, the mass and inertia term (left sides of the equations) must be negligible, which can be achieved by decreasing $m$ and $\mathbf{I}$ and increasing $\gamma_t$ and $\gamma_r$. Since we have chosen $D_{rot}$, we must set $\gamma_r = 1/D_{rot} = 10$, which gives us $\gamma_t = 3 \gamma_r = 30$, corresponding to the spherical case, as it was shown to be a good approximation for small aspect ratios\cite{Dijkstra19}. After checking the single-particle diffusion to ensure Brownian Dynamics were reached, the parameters of $m=0.1$ (per-particle mass), $I = 0.1 $, $\delta t = 0.001$ were found to be efficient. This then gives a Lennard-Jones equivalent timescale of $\tau' = \sqrt{0.1} \approx 0.361$. Aside from the particle density, we varied the magnetic interaction strength via the coupling parameter $\lambda = \frac{ \mu_0 \mu^2}{4 \pi k_B T \sigma^3} = \frac{\mu_0}{4 \pi} \mu^2$ with dipole moment $\mu = |\vec{\mu} |$, and the P\'eclet number $\text{P\'eclet} = v_0 \gamma_r$. On a technical level, the simulations were carried out using the simulation software package ESPResSo\cite{Weik2019,Shendruk16} version 4.2.2.. Since the simulations were started with random orientations and semi-random placement (each particle was placed on a random grid point, with grid spacing to avoid overlaps), there was an initial relaxation phase of $5 \cdot 10^4 \delta t$ before configurations were sampled. For systems with high values of $\lambda$ and low $Pe$ (such as $Pe=0$), these early configurations are not necessarily representative, as the magnetic interactions require a relaxation phase. To monitor this, we checked the dipolar energy contributions and excluded earlier timesteps at which this energy was significantly decreasing. For the $Pe=0$ comparison cases, we ran the simulation for an additional $8 \cdot 10^7 \delta t $ to ensure better equilibration. This does not mean that these cases are in the ground state, but it should ensure that the results shown are in a more representative distribution.

\begin{figure}
\centering
\includegraphics[scale=0.27]{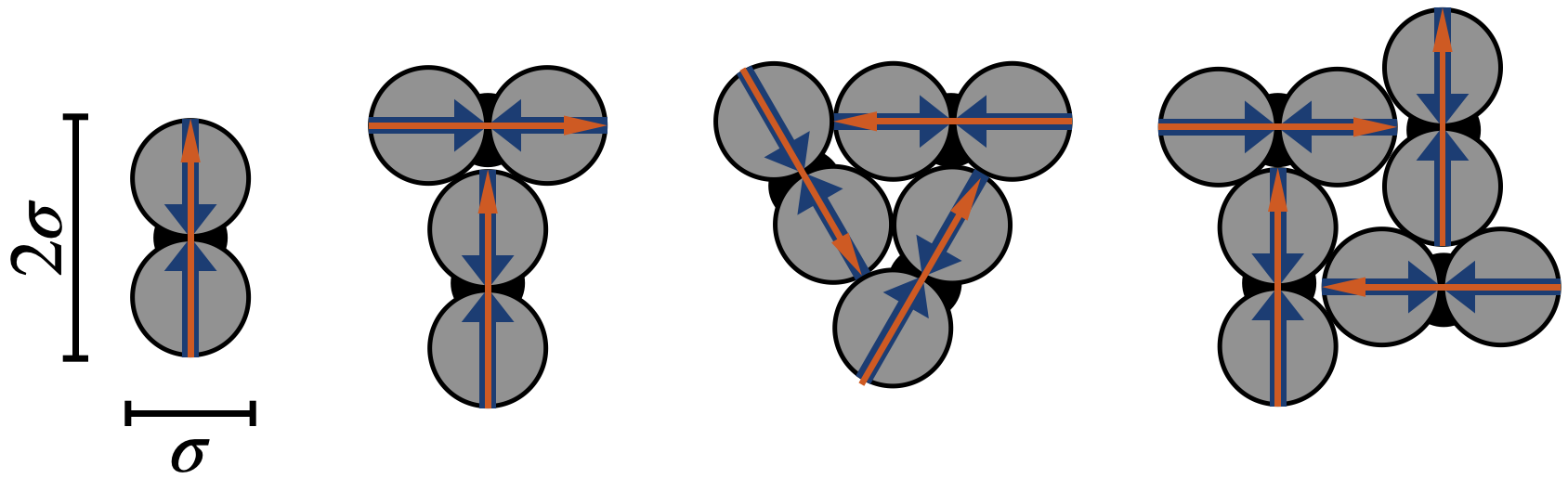}
\caption{\label{fig:simmodel} Left to right: one, two, three and four quadrupolar active particles in their respective ground states. gray denotes the steric repulsive particle shape, while the orientation of the two magnetic dipoles is shown in blue and the direction of the activity is shown in orange. To aid the eye in distinguishing which particles belong together, an additional black circle is added behind the connection point where the two dumbbell components meet.}
\end{figure}

\subsection{\label{subsec:model} Particle Model}

A sketch of two of the simulation particles used is shown in Figure ~\ref{fig:simmodel}. These were constructed using a rigid-body approach: each individual simulation particle consists of multiple interaction sites, but the translational and rotational equations of motion are only integrated for the particle center, from which the new positions of the sites are then derived. The particles consist of two steric repulsive spheres of diameter $\sigma$ (gray), which are modeled using the Weeks-Chandler-Anderson (WCA) repulsion\cite{Anderson71}, making the resulting particle apect ratio $2:1$. The WCA potential is given by:

\begin{equation}
        U_{W C A}(r)=\left\{\begin{array}{ll}{4\epsilon\left[\left(\frac{\sigma}{r}\right)^{12}-\left(\frac{\sigma}{r}\right)^{6}+\frac{1}{4}\right],} & {r \leq r_{c}} \\ {0} & {r \geq r_{c}}\end{array}\right.
        \label{eq:wca}
\end{equation}
\noindent where $r$  is the distance between the interaction site centres, with a cut-off distance $r_{c}=2^{1 / 6} \sigma$, $\epsilon = 1$ is the repulsion energy scale, which corresponds to the chosen energy unit $k_BT = 1$. Furthermore, each particle is active in the sense of the active Brownian particle model: a fixed self-propulsive force, with resulting velocity $v_0$, is applied along the director (orange), which is parallel to the long axis of the particle. To create the quadrupolar interaction, two magnetic dipole interaction sites are situated in the centers of each dumbbell component, with the magnetic dipole $\vec{\mu}$ pointing towards the particle center (blue). The dipolar interaction between two interaction sites $m$ and $n$ is calculated as:

 \begin{equation}
    U_{d d}(r_{mn},\vec{\mu}_m,\vec{\mu}_n)= \lambda \left[\frac{\left(\hat{\mu}_m \cdot \hat{\mu}_n \right)}{r_{mn}^{3}}-\frac{3\left(\hat{\mu}_n \cdot \vec{r}_{mn}\right)\left(\hat{\mu}_m \cdot \vec{r}_{mn}\right)}{r_{mn}^{5}}\right],
    \label{eq:dipdip1}
\end{equation}

where $\vec{r}_{mn}$ is the separation between two sites, $\hat{\mu}_m$ is the unit vector co-aligned with dipole moment on site $m$ and $\lambda$ is defined as in Subsection \ref{subsec:model}. Each dumbbell carries two such dipole moments, at a distance of $\sigma$. As will be explored in the following subsection, the net interaction results in a two-particle ground state as shown in Figure ~\ref{fig:simmodel}.

\section{\label{sec:results} Results and Discussion}

\subsection{\label{subsec:passive} Magnetic Quadrupoles}

\begin{figure}
\includegraphics[scale=0.35]{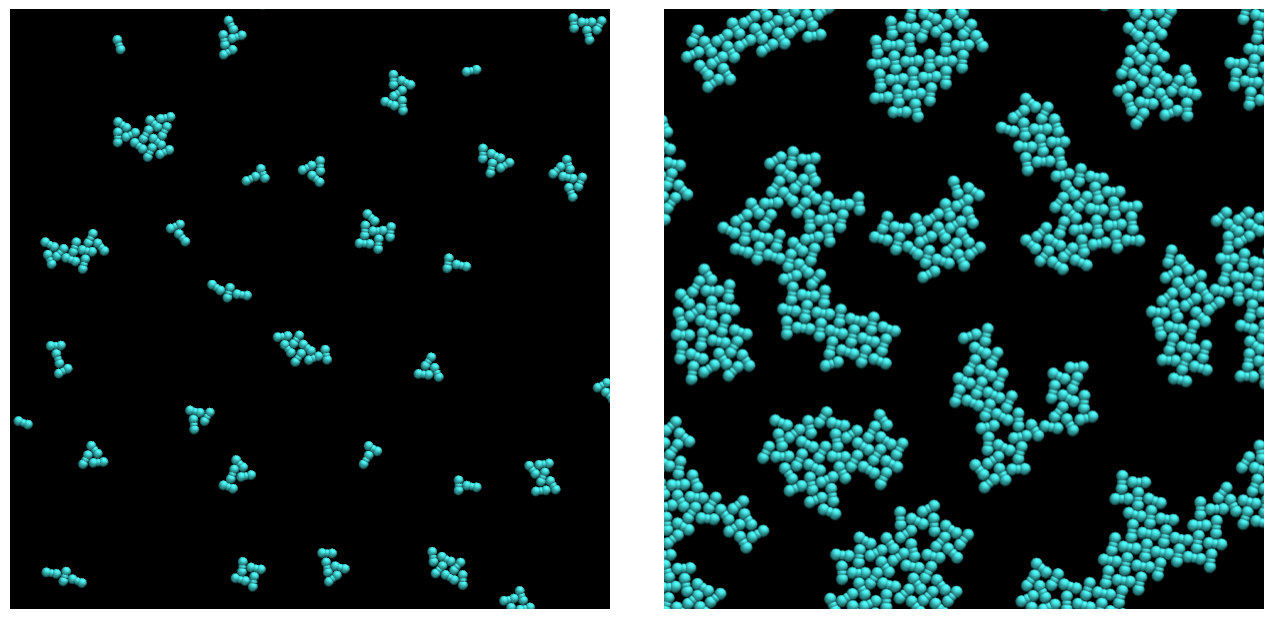}
\caption{\label{fig:snaps} Simulation snapshots from a quadrupolar dumbbell system with P\'eclet=0 and $\lambda =10$, for densities $\phi =0.05$ (left) and $\phi =0.3$ (right) after an initial relaxation phase. We see a variety of different interparticle alignments. As the system equilibrates, this aggregates combine into larger clusters with a predominately orthogonal base motif.}

\end{figure}

Before adding activity, a reasonable first question is to explore the microstructure we would expect from a passive magnetic system. Based on the configuration of the particles, we can easily see that the ground state for a $2$-particle system is orthogonal alignment. However, in multi-particle aggregates, it is no longer immediately obvious which configuration is most energetically favorable. To clarify this, we will proceed as follows. Starting with $N=2$, we check different particle configurations observed in simulations. We then calculate the cluster energy using only Equation ~\ref{eq:dipdip1}, without considering any thermal contributions ($T=0$) and assuming hard-spheres (no additional Weeks-Chandler-Anderson contribution). However, in order to match simulation, we do fix the closest-contact of two neighboring dumbbell components at $1.12 \sigma$. Starting with a pair of particles, the quadrupolar ground state is reached at orthogonal alignment (shown second from the left in Figure ~\ref{fig:simmodel}). For two particles, we have $4$ interaction sites. Assuming the upper particle to be centered at $(0,0)$, with magnetic moment parallel to the $x$-axis, the second particle would have position $(0,-1.5)$ with moment parallel to the $y$-axis. The actual dipole sites are then offset from the respective particle centers by $\pm 0.5$ along the long axes of the particles. This gives a total attraction of $U_{orth} \approx -1.56 \lambda$ at orthogonal alignment.In contrast, if we assume parallel alignment at the same distance, the repulsion is $U_{par} \approx 2.35 \lambda$. This clear difference might lead one to assume that particles will exclusively favor orthogonal alignments. However, if we look at the bulk simulations shown in Figure \ref{fig:snaps}, we see many different small clusters that do not appear to be orthogonally aligned. The reason for this can already be seen at $N=3$. Due to Euclidean geometry, we will not be able to align 3 particles at close-contact with orthogonal angles. At most, we can align 2 orthogonal pairs. The third interaction will be weaker due to the distance: if we for instance choose any triplet of particles of the four shown in Figure \ref{fig:simmodel}, we have a low attraction of $- 0.25 \lambda $, which gives a total energy of $\approx -3.37 \lambda $. While this is favorable, there is a superior option: if all three particles are brought into close-contact, forming a triangle (see Figure ~\ref{fig:simmodel}, second from the right). Then the particles are aligned at angle of $60^\circ$, and the attraction energy between each pair is $U_{tri} = -1.34 \lambda$, which yields a total cluster energy of  $\approx -4.02 \lambda$. This means that even though the pairwise interaction has a minimum at the orthogonal alignment, for three particles, the ground state is this triangular motif.

 Increasing the particle number to four, we are able to form a lattice motif as shown on the right side of Figure \ref{fig:simmodel}, which with 4 orthogonal pairs and two more distance neighbor interactions has the total energy of $\approx -6.74 \lambda$. This makes it more energetically favorable than any possible combination of a triangle and a single particle. These calculations can be continued for increasing numbers of particles, but the relevant point for this work is as follows: although the orthogonal lattice motif is the most energetically favorable, at finite temperatures and low $N$, there are multiple competing motifs which occupy local minima This ambiguity is also seen in the simulation snapshots, as shown in Figure ~\ref{fig:snaps}. At low densities, we see multiple different options for particle aggregates, while for higher densities, the orthogonal lattice is favorable.

 Although these two snapshots provide an intuitive qualitative insight, this does not provide the full picture of the relevant ground states. For this, we refer to an previous work on (electric) ellipsoidal quadrupolar$^{57}$ particles which discusses the strong dependence of the ground state on the particle shape. As our particles artificially stabilizes the orthogonal minimum, we recover the orthogonal lattice ground state discussed for the more symmetric particles instead of the herringbone structure found for more elongated particles. Interestingly, the authors also found a trihexagonal pseudo-lattice tiling featuring $60^\circ$ alignment between particles, which seems to be the most topologically similar to our triangle motifs. It would be interesting to extend this discussion to lower densities, including a variation of shapes, but unfortunately this is beyond the scope of our paper.

\subsection{\label{subsec:phase} Phase Behavior}

\begin{figure*}
\includegraphics[scale=0.54]{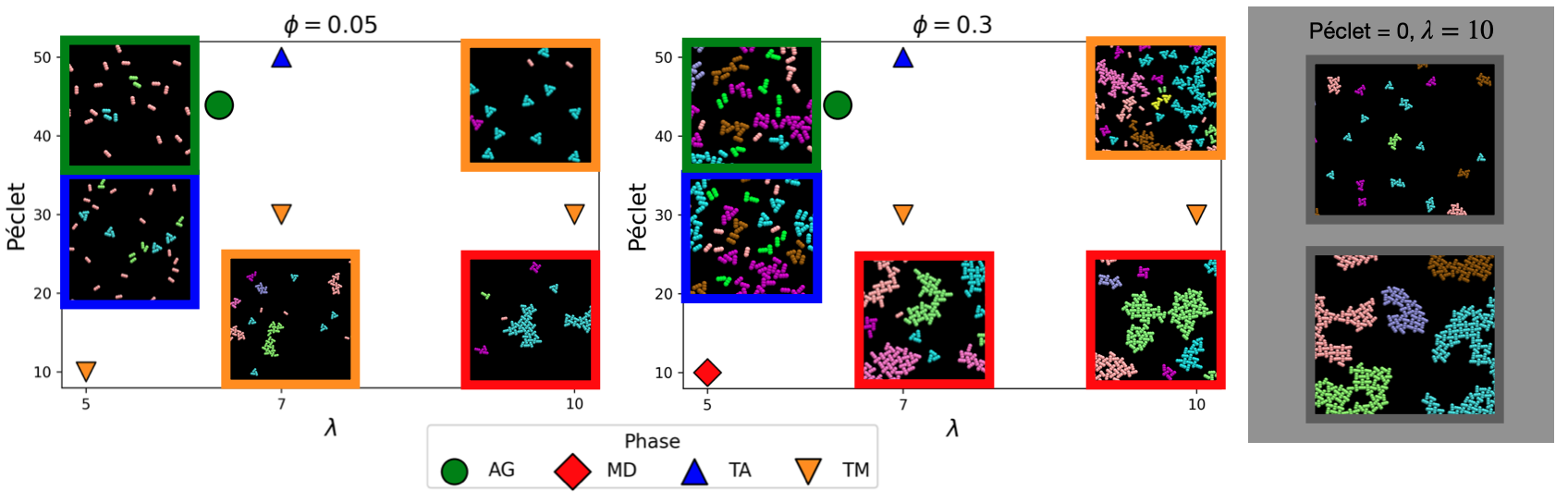}
\caption{\label{fig:phase} The state diagrams of active quadrupolar dumbbells, in dependence of the magnetic coupling constant $\lambda$ and the P\'eclet number at different area fractions $\phi$. Each symbol represents a different region in the state diagram: representative simulation snapshots are inset at a few key values, in which cases, the frame color denotes the state. We distinguish between 4 key states. Active-gaseous, in green, shows little aggregation except at high densities and no characteristic structures or ordering. Magnetically dominated, in red, shows aggregation typical of magnetic systems, albeit with what seems to be more orthogonal alignment between particles than one would expect for dipolar particles. At intermediate values of $\lambda$ and the P\'eclet number, we find states with more triangular motifs: either triangular-active (blue), which most closely resembles the active-gaseous system except for a preponderance of triangles, and triangular-magnetic (orange), which similarly features a more magnetic aggregation pattern, except for a peak in aggregation at $n=3$. The dumbbells inside these snapshots are colored based on the number of particles in their aggregate, with a coloring pattern that repeats for large $n$, shown with an additional fictitious connecting particle in the middle as a visual aid. It should be noted that the snapshots cannot fully represent the system due to size constraints, and many of the clusters shown are truncated. In the gray boy (right), we show larger snapshots from the  P\'e=0 case, for $\phi = 0.05$ (top) and $\phi = 0.3$ (bottom).}
\end{figure*}

To understand the impact of activity on the system, and to see if the quadrupole-induced orthogonal alignments occur in the bulk, we computed the phase diagram shown in 
Figure ~\ref{fig:phase}. From even a brief visual inspection of the simulation snapshots shown, it is evident that the phase behavior of the system does not map neatly onto either the magnetic or active phase diagrams. Most surprisingly, even at a brief visual inspection, some densities seemed to feature an excessively high amount of 3-particle triangular aggregates. As discussed in the previous subsection, while this is the ground state for this particle number, an even larger aggregate would be even more energetically favorable. From studies done on the clustering of passive magnetic particles\cite{Russo11,Sciortino12}, and our calculations, we do not expect any specific cluster sizes to be preferred in the aggregation process. However, the pervasiveness of these triangles is borne out by statistical analysis (as will be discussed in detail in the following section). To therefore characterize the emergent states, we devised a metric based on the characteristics of $p(n)$, which describes percentage of particles in a cluster of size $n$. In order to properly consider the possibility that particles form less magnetically favorable clusters in regions of high-activity, clustering was determined using a purely distance-based criterion. While there are no sharp phase transitions in the system, we nevertheless find four distinct states which we can characterize as follows. Firstly, for low values of the magnetic coupling constant $\lambda$, and high P\'eclet, the aggregation behaves more in line with a non-magnetic active system. Clusters do form with increasing density, but there are no preferred motifs, and smaller aggregates predominate. We denote this as "AG" (active gaseous, green). At the opposite limit case of high $\lambda$ and low P\'eclet, we use "MD" to denote the magnetically dominated case (red): here, we see the trend toward aggregation specifically into larger clusters that we would expect from magnetic systems at such values of the coupling constant and density. An initial visual inspection does suggest a visual difference from dipolar magnetic systems in that the clusters show a grid-like structure based on orthogonal alignment, but by no means exclusively. This comparison is made explicit by the P\'eclet=0 case shown in the gray box (right), where we see a distribution of clusters of various sizes at low density and a strongly aggregated phase with orthogonal alignment at higher densities.

This leaves the majority of the phase diagram, which exhibits the interesting new state mentioned before: that of active triangular clusters. We further subdivide this state into two distinct states. The "TA" (triangular active, blue) is characterized by a pattern (or lack) of aggregation that would be expected from a non-magnetic active system, but with a peak at $n=3$ that appears to give rise to triangles. Similarly, "TM" (triangle magnetic, orange) denotes a system with a baseline aggregation consistent with what would be expected from a magnetic system, but with a significant peak at $n=3$. With increasing density, these differences are not possible to distinguish visually, which is why the cluster distribution function $p(n)$ was chosen. Aside from the direct balance between the activity and magnetic interaction strength, the density $\phi$ of particles also plays an important role: as $\phi$ increases, the magnetically dominated fraction of the phase diagram grows, although the active and triangular-active states are maintained. This suggests that the microstructure of the system can be tuned based on the balance of activity to magnetic coupling, while remaining robust across reasonable densities, as $\phi = 0.3$ is already considered a very high density for a magnetic system.

In the electronic supplementary information (ESI), we have additionally attached short renderings of the lower-density simulations where an additional key aspect of the system is highlighted: the spinning of the aggregates. This is most notable for the small aggregates: in particular, we can see that the triangular aggregates and the less-frequent four-particle aggregates spin rapidly. These visually resemble classic windmills (or modern wind-turbines) after which the motion is named. On an ensemble level, we see that, as expected from the lack of chirality, there is no overall rotational ordering of the fluid. 

\begin{figure}
\includegraphics[scale=0.25]{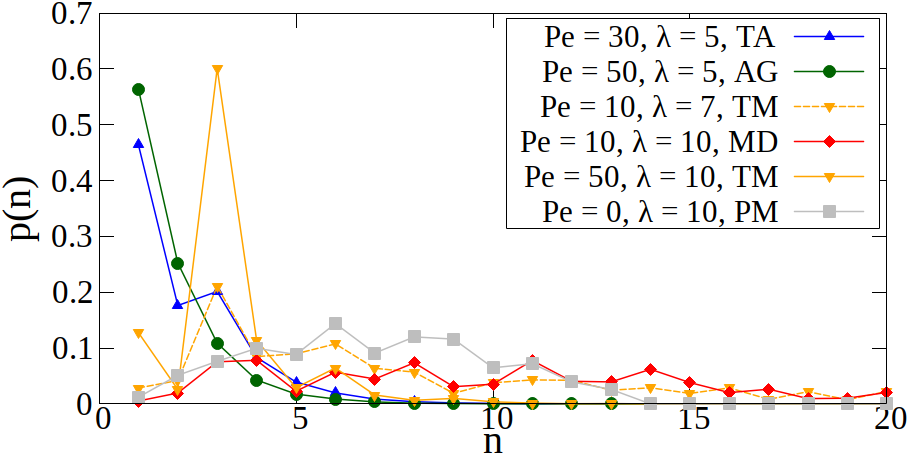}\\
\includegraphics[scale=0.29]{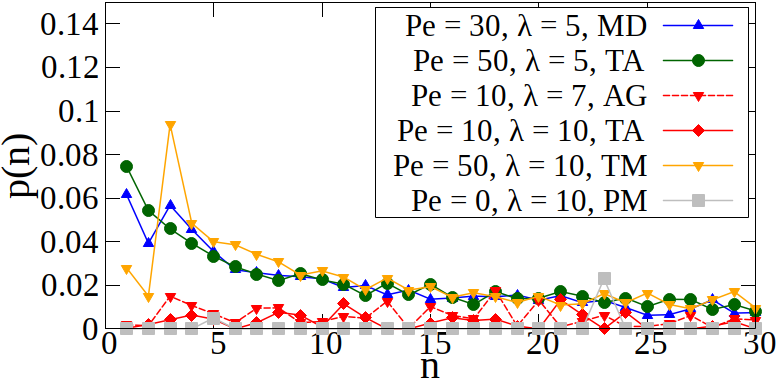}
\caption{\label{fig:cllowdens} The percentage of particles in a cluster of size $n$, truncated at $n=20$ and $n=30$ in order to highlight the $n=3$ behavior. Colors also correspond to the states indicated in Figure ~\ref{fig:phase}, with the P\'e=0 system shown in gray as "PM" (pure magnetic). Top: $\phi = 0.05$, Bottom: $\phi =0.3$. Note the difference in $p(n)$ axis, as at high densities, many particles are contained within less frequent, larger clusters. }
\end{figure}

For a quantitative, more precise exploration of the microstructure, we turn to Fig.~\ref{fig:cllowdens}, which shows the cluster distribution function $p(n)$, truncated at the size of 20 for aggregates. This plot illustrates the specific state designations, chosen at the same characteristic densities highlighted by the snapshots in Figure ~\ref{fig:phase}. Beginning with the $\phi = 0.05$ low-density plot, we see that all but the active-gaseous (green), magnetic-dominated (red) and pure magnetic (gray) case have a peak at $n=3$, which corresponds to the visually observed triangles. While the distinction is less clear from simulation snapshots, here we can observe that the triangular-active curve (blue) does qualitatively follow the active-gaseous, albeit with more overall aggregation. This stands in contrast to the two triangular-magnetic curves (orange), which qualitatively exhibit the same baseline of resembling the magnetic cluster distribution, but with a peak at $n=3$. These two curves also serve to illustrate that there is some variety in the TM state, due to the underlying variance in the magnetic behavior. While the activity can reduce the aggregation, we would expect a transition away from a stable cluster-fluid past a percolation threshold as $\lambda$ and $\phi$ increase. In practical terms relevant to this plot, we would expect for lower $\phi$ and $\lambda$ to still find a variety of smaller aggregates growing in size (for the dipolar case, following Wertheim theory\cite{wertheim84,wertheim86}), while at higher values, a curve similar to the one shown for the magnetically dominated and pure magnetic cases is expected (few, very large aggregates). It appears that the activity is impeding the magnetic aggregation such that we can still access part of this state, despite the magnetic interaction being so high (see Section ~\ref{subsec:passive}) that this is not expected. 

In the higher density plot, we see that the increasing aggregation has diluted the respective peak heights. It must be noted that choosing a plot range which will clearly show $n=3$ to illustrate the state criterion necessitated some truncation: at these densities, most particles are aggregated in large, but therefore statistically less frequent clusters, resulting in a mostly flat distribution. However, we can still see a distinction between the different state that is no longer visible in the simulation snapshots. Moreover, we see that with increasing density, the high-$\lambda$ and low P\'eclet triangular magnetic cases have collapsed into the magnetically dominated case, which agrees with the pure magnetic case, as seen in the phase diagram. This means that now the internal structure of the clusters is a more relevant question.

\begin{figure}
\includegraphics[scale=0.35]{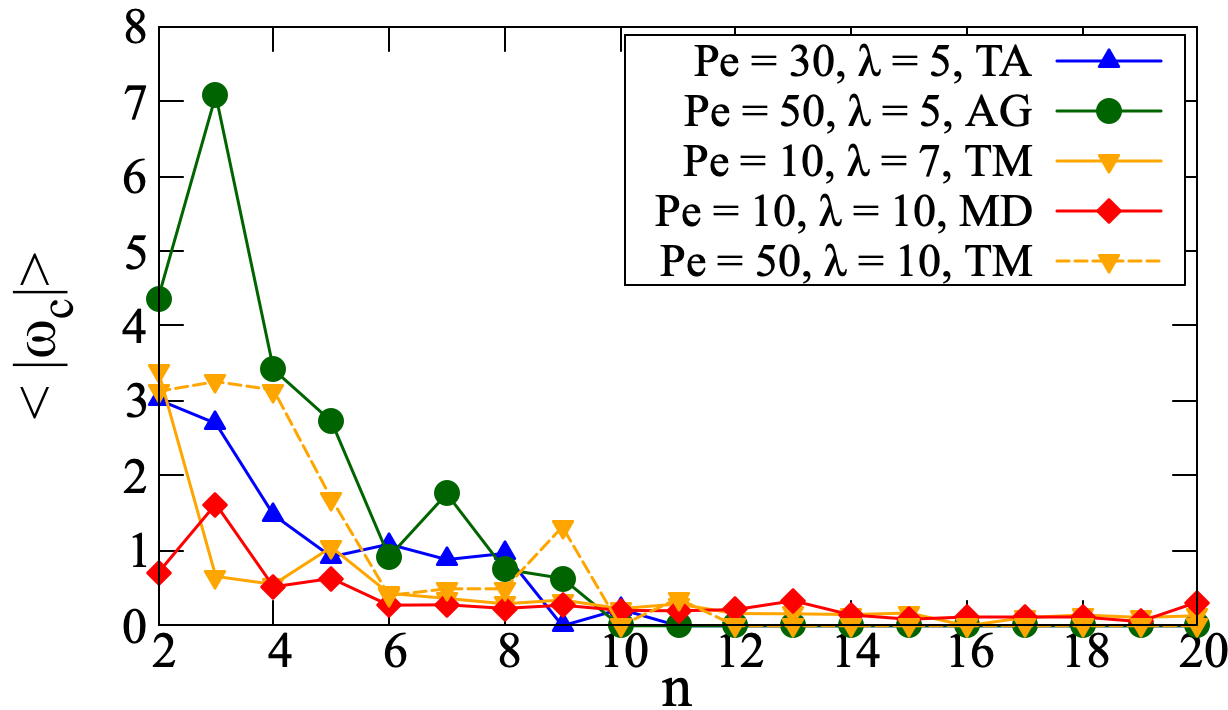}
\caption{\label{fig:averageavel} Average angular velocities of clusters of a given size at density $\phi = 0.05$. Aside from the P{é}clet number, the speed of the aggregates depends non-trivially on the number of particles. }
\end{figure}

Now that we have characterized these different cluster sizes, we briefly turn to characterizing their windmilling rotation by computing the average angular velocities of clusters of a given size. This is calculated by computing $\vec{\omega}^{(n)} = (\sum_{i=1}^n (\vec{r}_i \times \vec{v})/|\vec{r}_i|^2)/n$ for each cluster of size $n$, where $\vec{v}_i$ is the particle velocity minus the velocity of the center of mass and $\vec{r}_i$ is the vector that goes from the center of mass to the dumbbell center. As our particles may only rotate around the $z$-axis, there are two possible signs of $\vec{\omega}^{(n)}$. Thus, in Figure \ref{fig:averageavel}, we average over the absolute value of each cluster of size $n$. The strongest dependency is on the P{é}clet number, which determines the range of the average cluster angular velocities. Interestingly, the speed of pairs is the most stable across different systems, presumeably because of its lack of variation in topology. Similarly, three-particle aggregates  windmill at least as fast as other larger aggregates, although the actual velocity depends on their environment. In the more activity-dominated systems (blue, green) there is a significant drop-off in cluster speed past three-particles. Contrasting this, the high-P{é}clet triangular active state (solid, orange) shows an almost equal rotation speed for any of small aggregates from two to four particles. Interestingly enough, while the average speed of the three-particle aggregates found for P{é}$=10$, the triangular magnetic system (orange dashed) has a lower average cluster speed for three-particle aggregates than the magnetically dominated system (red). This is possibly related to issue of cluster fusion, and fission, as well as mutual collisions with other clusters which provide a fluctuating background, slowing down the rotation compared to the more isolated triangular spinners in the magnetically dominated system composed of less-mobile, large aggregates. For the larger cluster sizes, we see the average rotation rapidly decrease to close to zero. Although these do rotate slightly due to asymmetry, the speed is on average very low.

\subsection{\label{subsec:micro} Cluster Microstructure}

\begin{figure}
\includegraphics[scale=0.22]{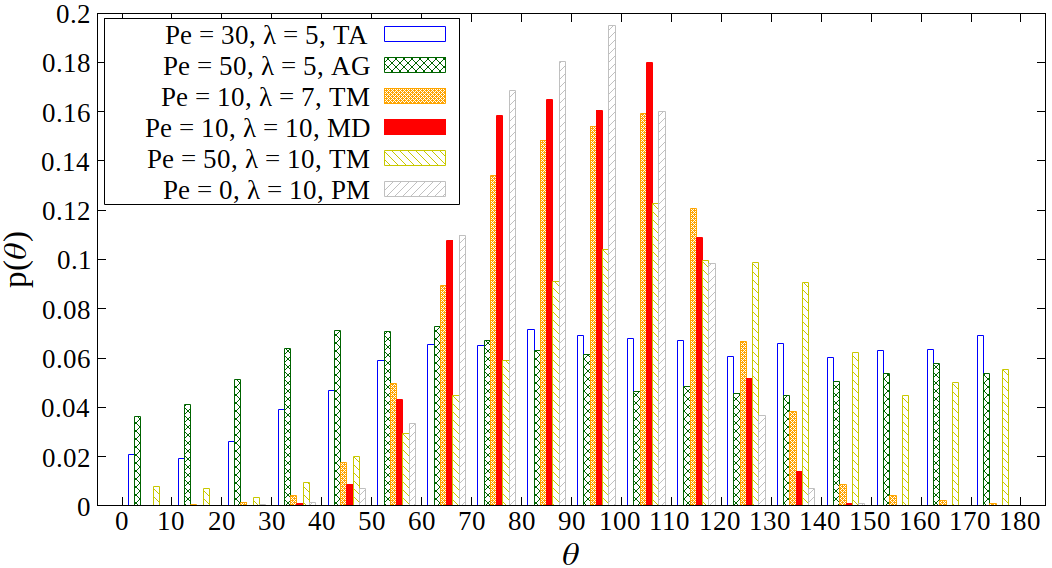}\\
\includegraphics[scale=0.22]{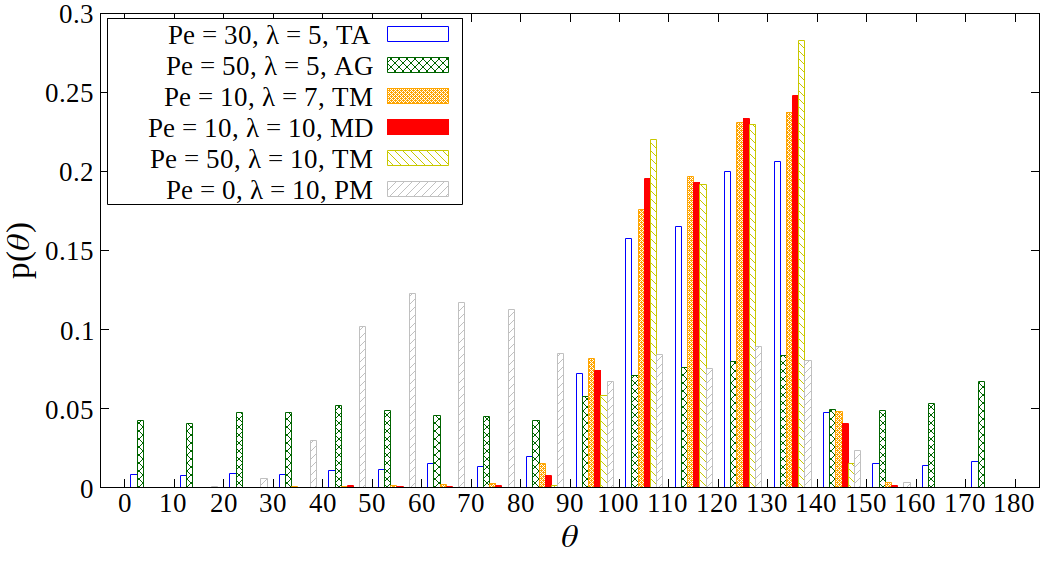}\\
\includegraphics[scale=0.22]{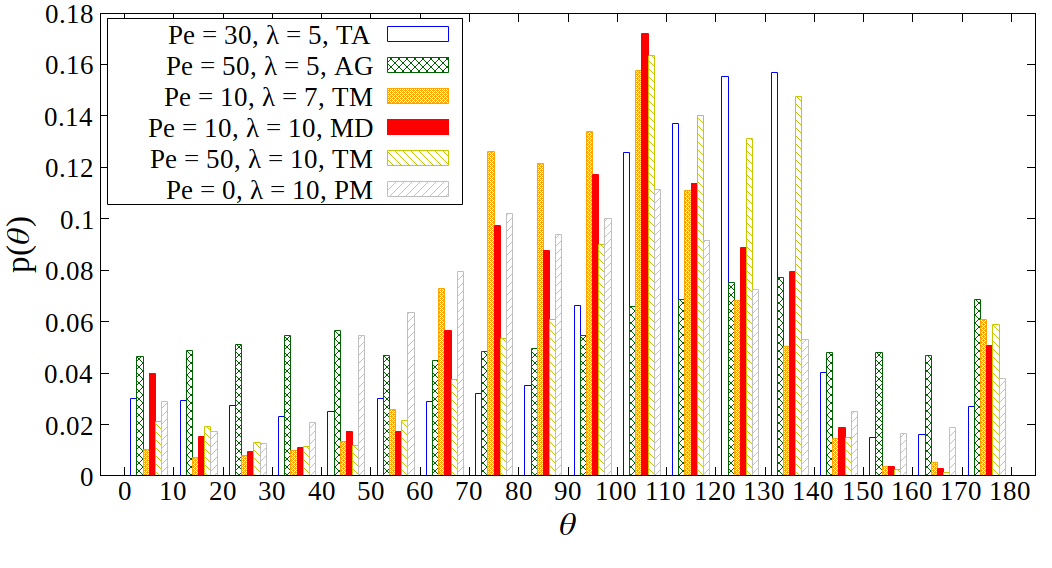}
\caption{\label{fig:angles} Histograms showing the count $p(\theta)$ of angles $\theta$ occurring within a cluster of size 2 (top), 3 (middle) and 4 (lowest), normalized by the count of angles. For pairs, we see characteristic angles close to 90, and a symmetry with respect to 90 degrees. For the triangles, we see the combined effect of activity in the break of symmetry, although much less so for the magnetically dominated systems and not in the purely magnetic system. Finally, for the larger aggregates, we see the effect of increasing magnetic strength. While the aggregation distribution of triangle-magnetic at high P\'eclet follows more magnetic patterns, the internal structuring of the clusters is still altered by the activity.}
\end{figure}

Until this point, we have often presumptively referred to $n=3$ particle aggregates in specific systems as "triangles". Our definition of clustering also currently conflates energetically very different structures, since the only characterizations are based on the number of particles. This is interesting to explore since there does appear to be a visual difference in microstructure. As discussed in Section ~\ref{subsec:passive}, the magnetic minimum is orthogonal and a purely active particle would not have a specific preferred alignment, aside from shape-anisotropy effects: in other words, there is no unique direct "triangle-inducing" parameter. To gain an understanding of the micro-structural motifs, we investigate the mutual angles of neighboring particles in a cluster of size $n$ as shown in Figure ~\ref{fig:angles} for $\phi = 0.05$. It should be noted that, although the magnetic quadrupolar interactions are symmetric along the long axis of the particle, the activity is polar: therefore, angles are shown from $0^\circ$ to $180^\circ$. A schematic defining this angle is shown in the Supplementary Information.

Beginning with the topmost plot, which shows the angle distribution of the pairs, we see that for both the active gas (green) and the triangular active (blue), there are practically no preferred angles. However, it must be noted that these plots are all dependent on the number of clusters of size $n$ in the system to obtain statistics, as these directly depend on how many angles are averaged over. The initial triangular magnetic and magnetically dominated systems show a roughly symmetric distribution around $90^\circ$, suggesting the formation of roughly orthogonal pairs. These two distributions also show few alignments close to $0^\circ$ or $180^\circ$ degrees, which are the most unfavorable in terms of the quadrupole interactions. This symmetric distribution around $90^{\circ}$ is expected based on the quadrupolar interactions, especially for large values of $\lambda$. Yet, when we consider the high-P\'eclet, high-$\lambda$ triangle fluid, angles above $90^\circ$ degrees are preferred and additionally, some pairs close to $180^\circ$ degrees are found. It is the polarity of the particles which breaks the symmetry. This effect is slightly counterintuitive, but can be explained when considering that there are two possible angles for each simple geometric alignment depending on the orientation of the director. Close to the magnetic minimum of 90 degrees, this appears to be irrelevant. But at magnetically unfavorable alignment, the higher angle would correspond to the particle activity pointing inwards toward the other particle. This would suggest a sort of jamming effect: for magnetically unfavorable angles, particles with co-aligned activity can separate due to the torque of the magnetic repulsion. However, if the particle is pointed in the opposite orientation, it is not able to separate. 

For three-particle aggregates, shown in the middle plot, we see a pattern of peaks centered around $120^\circ$ degrees, suggesting the formation of triangles where all of the particles are aligned to push against their neighbors. These are consistent whenever such clusters are abundant- notably, they are present in the triangular active case (blue) which did not show a preference for pairwise aggregates, but not found in the active gaseous case (green). This is also the case where we see the most pronounced effect of the activity, as the distribution of angles for the purely magnetic (gray) case is clearly symmetric. As can also be seen in the supplementary information videos, active triangles necessitate a specific orientation of the polarity to remain stable: the direction of the activity needs to point inwards. This leads to the angle being counted as $120^{\circ}$, as described in more detail in the Supplemental Information. In the case where the activity is set to zero, the particles are no longer polar and this distinction becomes irrelevant, which is why we see the distribution does not have this strong symmetry break.

For the four-particle aggregates, the behavior now strongly diverges. The active gaseous state still forms a consistent baseline (green), while the triangular active case (blue) shows a similar motif to the $n=3$ particle aggregates, suggesting hinting at the continued stability of the triangular base motif, perhaps with added particles. This also holds for the triangular magnetic case with high activity (yellow), further suggesting that the activity drives the formation of triangular motifs. In contrast, the two remaining systems (red, orange), both with higher $\lambda$, have a cluster distribution that is now shifted to include values around $90^\circ$: more resembling a superposition of the $n=2$ orthogonal and $n=3$ triangular alignments. Interestingly, there is a slight peak around $0^\circ$ and $180^\circ$. This could either be from jamming as conjectured for the pair case, or some attachment of an additional particle to an existing triangular motif.

\begin{figure}
\includegraphics[scale=0.21]{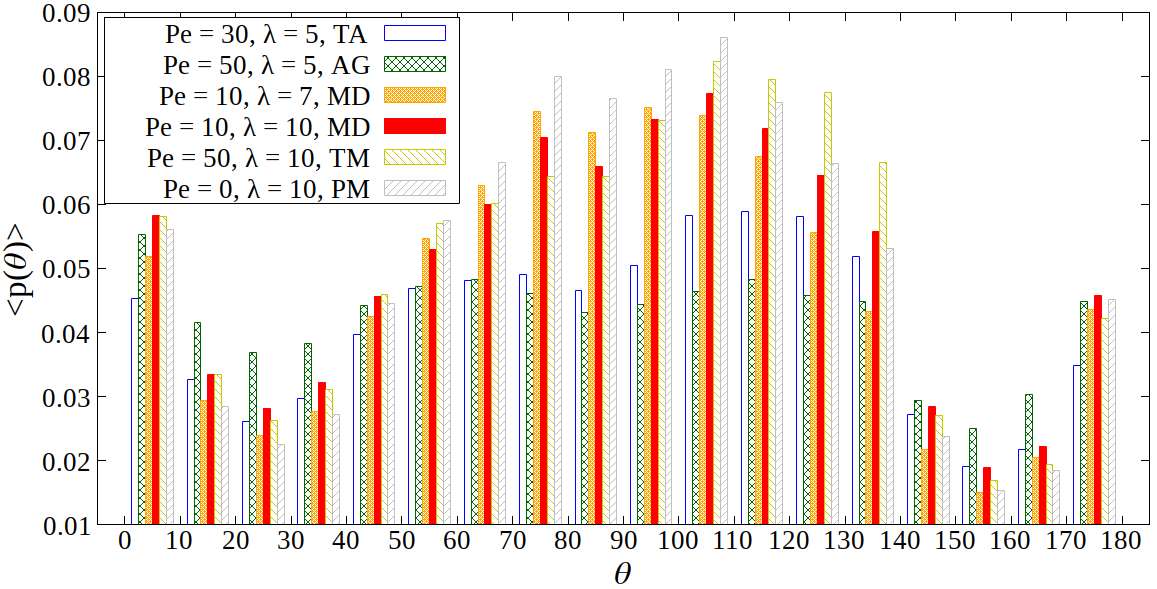}
\caption{\label{fig:denseangles} Histograms showing the average count $ \langle p(\theta) \rangle$ of angles $\theta$ occurring within a cluster of sizes between $2$ and $200$, at $\phi = 0.3$. Overall, we see that the more activity-dominated systems show a lower preference for specific angles, and a reduction of the symmetry break seen in the low-density systems.}
\end{figure}

To understand how much of this distinct patterning in preserved in larger clusters, we turn to the denser systems of $\phi = 0.3$. However, as we have seen in Fig.~\ref{fig:cllowdens}, these systems feature overall fewer clusters, meaning that choosing a specific cluster size $n$ would not necessarily yield good statistics. We therefore average over all clusters with sizes between $2-200$ particles, the results of which are shown in Fig.~\ref{fig:denseangles}. Both of the more activity-dominated systems (blue, green) show a more equal distribution of angles. Interestingly, the triangular active system is now almost symmetric around $90^\circ$. This suggests that some triangular-type motif remains, even as the polarity becomes less important in systems where the particles are less able to move freely.  The magnetically dominated systems (orange, red) and the pure magnetic (gray) show a similar distribution as already found in the $4$-particle structures of the low-density suspension, allowing for both orthogonal and triangular-based motifs, albeit now with significant peaks for neighboring particles at the angles around $0^\circ$ and $180^\circ$. The triangular magnetic system shows a similar spread of angles, but higher values around $110^\circ - 140^\circ$ similar to the triangular active case, suggesting at least some continued presence of the triangular motif. This means that despite the averaging, and certain visual similarities, the microstructure still exhibit different underlying trends.

To recapitulate, we have now seen that the clustering behavior is tuneable based on the relative strengths of magnetism and activity. While this clustering distribution can be chosen, the internal structure can also be tuned to more reflect triangular or orthogonal structures, although there is no strict phase transition.

\section{\label{sec:concl} Conclusions}

Our work here introduces a novel system of active magnetic quadrupolar dumbbells. Despite these building blocks having individually known behavior, the combination creates an entirely new set of states. The quadrupolar magnetic aggregations lead to an orthogonal alignment in the magnetic limiting cases, which is destabilized by increasing activity. The most novel states is a suspension of windmilling triangular aggregates, which can be found in a more dilute form as a form of increased aggregation in an otherwise active bath, or as an unexpected peak in otherwise magnetic suspensions. Even if the magnetic interactions are allowed to dominate, there is still a significant deviation from the purely orthogonal ground state. The microstructure of these aggregates can be tuned by the balance between the activity and quadrupolar interaction. Despite the lack of chirality, the particles form spinning aggregates: like windmills, these can be based on the triangular three-particle alignment or an orthogonal based four particle grid. Due to the simplicity of the system, it could easily be implemented experimentally, for instance in systems of active granular particles, and allows for enhancements such as using an external magnetic field to direct the particles. Moreover, it would be possible to alter the alignment of the dipolar sites: either creating quadrupoles aligned perpendicular to the activity, or intermediate potentials which are neither quadrupolar nor dipolar. Another promising avenue for future study would be to investigate a three-dimensional systems. For more dense systems, Gay-Berne particles with an electric quadrupole moment have been found to exhibit liquid crystalline phases\cite{BATES01021998,Hansen97}. Similar to systems of Laponite or Gibbsite platelets, we would also expect to see "house-of-cards" type structures as a generalization of the herringbone- and triangular motifs to three dimensions. Once we also consider the effects of activity on these structures, we would expect to see helical motion\cite{Witten2020}, as the two-dimensional windmilling rotations would translate into three-dimensional chirality. Overall, this work seeks to illustrate an active system with tuneable microstructure that gives rise to many new lines of investigation. 

\section{\label{sec:suppl} Supplemental Material}
In addition to the main manuscript, a more detailed description of the angle definition (including schematics) and short video clips depicting trajectories of the system are available online. 

\begin{acknowledgments}
We wish to acknowledge the support of DFG project LO 418 / 29.
\end{acknowledgments}

\section*{Data Availability Statement}

The data that support the findings of this study are available from the corresponding author upon reasonable request.

\bibliography{aipsamp}

\end{document}


\section{Supplementary Information}

\subsection{Details of the interparticle angular alignment}

\begin{figure}[h]
\centering
\includegraphics[scale=0.35]{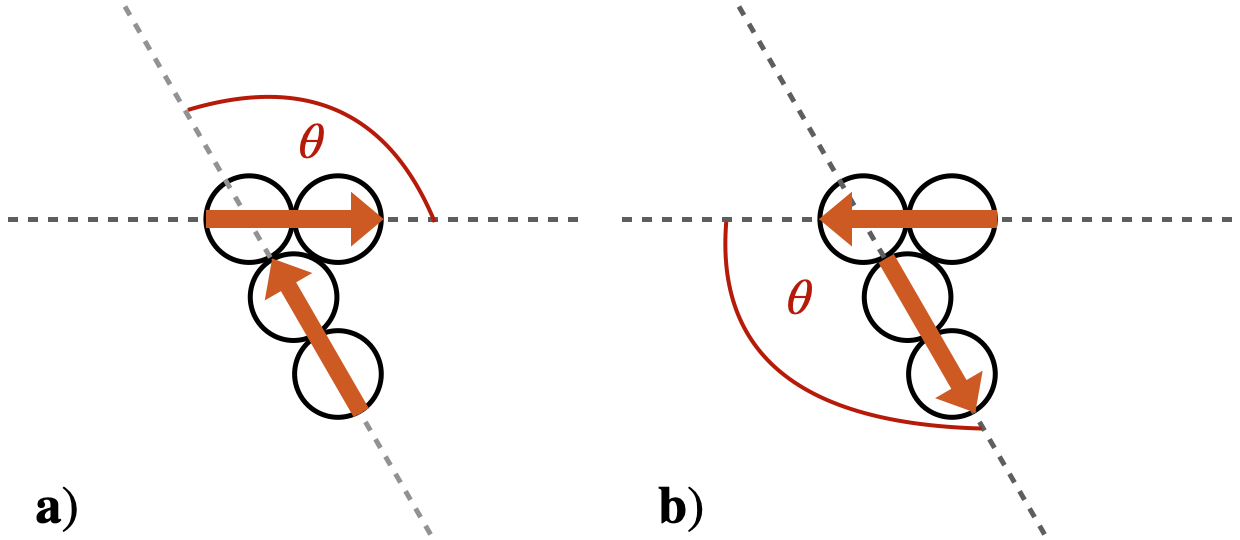}
\includegraphics[scale=0.35]{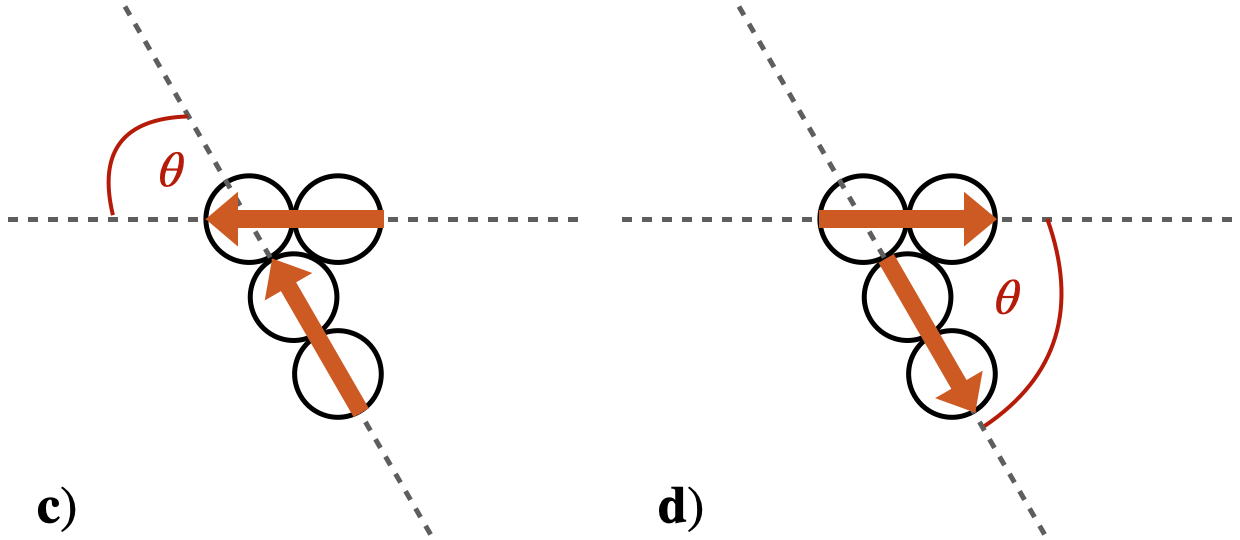}
\caption{\label{fig:pairangle} Schematic showing how the angle $\theta$ between particle directors is defined; the particle director is shown in orange, with long axis of the particle shown as a grey dashed line. Despite identical positions, the different particle directors mean that there are four possibilities, and two possible angles. In this specific case, the configurations a) and b) have an angle of $\theta = 120^{\circ}$. The configurations c) and d) have an angle of $\theta = 60^{\circ}$. Although these aggregates have the same attractive quadrupolar energy, simulations with with P\'e $>0$, almost exclusively find alignment a) in triangular spinner clusters. For pairs, both a) and c) appear to be equally frequent.}
\end{figure}

In Section III.C., the cluster microstructure is discussed based on the interparticle angle $\theta$. This angle is defined as the angle between the two particle directors $\hat{n}_i$ and $\hat{n}_j$ for particles $i$ and $j$; more specifically $\theta = \arccos{(\hat{n}_i \cdot \hat{n}_j)}$, as directors are unit vectors. This is not the same as the mathematical definition of an angle between two planes, as depending on the orientation of the particle directors, either the smaller or larger angle between the long axes of the particle is chosen. This is illustrated in Figure \ref{fig:pairangle}: despite the physical position of the particles being identical, since the directors can point in either direction along the long axis of the particle, there are actually four different possible configurations, with two possible angles. This is intentional, as these directors represent the direction of the particles' self-propulsion. 

